\documentclass[9pt,twocolumn,twoside]{opticajnl}
\journal{opticajournal} 

\setboolean{shortarticle}{true}

\newcommand{\fa}{{\it Fiber~\textbf{A}}}
\newcommand{\fb}{{\it Fiber~\textbf{B}}}

\usepackage{lineno}

\title{Addressing modulational instability in anti-resonant hollow-core fibers for pulse compression}

\author[1,2]{Michael Hemsworth}
\author[1,2]{Arthur K. Mills}
\author[3]{TJ Hammond}
\author[1,2,*]{David J. Jones}

\affil[1]{Department of Physics and Astronomy, University of British Columbia, Vancouver, British Columbia, V6T 1Z1 Canada}
\affil[2]{Quantum Matter Institute, University of British Columbia, Vancouver, British Columbia, V6T 1Z4 Canada}
\affil[3]{Department of Physics,University of Windsor, Windsor, Ontario N9B 3P4}

\affil[*]{djjones@physics.ubc.ca}

\begin{abstract}
When pulses propagate in gas-filled anti-resonant hollow-core fibers (AR-HCFs) modulational instability (MI) can lead to pulse break-up and loss of coherence. In pulse broadening and compression schemes, MI is a parasitic effect that induces significant shot-to-shot fluctuations of the peak power of compressed pulses and increases rapidly over a narrow range of input pulse energies. In this work we use experimental studies and supporting numerical simulations to compare two AR-HCFs that are chosen to enhance or suppress MI. We demonstrate that judicious selection of the wall thickness of the anti-resonant elements (AREs) can drastically reduce the MI gain, thereby increasing the limit of pulse energy scaling of stable ultrafast pulse compression.
\end{abstract}

\setboolean{displaycopyright}{false} 

\begin{document}

\maketitle

Gas-filled anti-resonant hollow-core fibers (AR-HCFs) provide a useful platform for several ultrafast nonlinear processes, enabled by long light–matter interaction lengths and precise dispersion control. They have been applied in molecular spectroscopy \cite{kotsina_ultrafast_2019}, dispersive wave emission \cite{joly_bright_2011,mak_tunable_2013}, four-wave mixing \cite{belli_highly_2019,couch_ultrafast_2020}, and pulse compression \cite{kottig_efficient_2020, schade_scaling_2021}. Compared to conventional hollow-core fibers, AR-HCFs have several advantages, including a smaller footprint and substantially lower loss for reduced core sizes \cite{russell_hollow-core_2014,wei_negative_2017}. These advantages make them particularly attractive for compressing relatively low-energy (\textmu{}J-level) pulses. However, AR-HCFs exhibit spectral anti-crossings between core and cladding modes that introduce narrow loss bands and sharp dispersion variations \cite{tani_effect_2018}, which can give rise to modulational instability (MI) \cite{agrawal_nonlinear_2019}.

MI is a four-wave mixing process phase-matched by the combined action of dispersion and the Kerr nonlinearity \cite{agrawal_nonlinear_2019}. Seeded by zero-point-fluctuations, MI leads to the growth of spectral sidebands, which strongly modulate the pump pulse and can eventually lead to pulse break-up. As has been previously shown, the high sensitivity to initial pulse parameters produces significant pulse-to-pulse power fluctuations that can, in turn, lead to loss of pulse train coherence \cite{dudley_coherence_2002}. Since MI tends to occur for longer input pulse durations \cite{dudley_supercontinuum_2006,schade_scaling_2021}, it becomes particularly important to quantify for applications such as nonlinear pulse compression.

Here, we build on the initial theoretical work of Köttig et al.~\cite{kottig_modulational-instability-free_2020} and experimentally confirm the conditions and effects of both enhancement and suppression of MI during self-phase modulation (SPM) broadening in AR-HCFs. Through experiments and numerical simulations, we investigate the performance of two commercially available AR-HCFs, one that strongly supports MI and one that suppresses it. We investigate the energy scaling of spectral broadening and show that the exponential growth of MI sidebands rapidly limits practical spectral broadening because of imparted pulse-to-pulse peak power fluctuations. The unstable nature of the MI sidebands is experimentally verified using single-shot dispersive Fourier transform (DFT) measurements. We confirm strong agreement between experimental results and numerical simulations, where the latter show that even modest levels of MI result in significant peak power fluctuations after pulse compression. Finally, we demonstrate the input pulse energy and gas pressure scaling of these MI-induced fluctuations on the output pulse peak power stability.

The anti-crossing features in the AR-HCFs arise when the core and cladding modes become strongly coupled due to resonances within the walls of the microstructured cladding, or antiresonant elements (AREs) \cite{kottig_modulational-instability-free_2020}. The central wavelength of the $m$th resonance can be approximated by $\lambda(T) = (2T/m)\sqrt{n^2 - 1},$
where $T$ is the ARE wall thickness and $n$ is the cladding refractive index. Near these resonances, loss increases dramatically \cite{archambault_loss_1993} and group-velocity dispersion changes rapidly, producing the complex dispersion profiles characteristic of AR-HCFs \cite{tani_effect_2018}. As MI gain is directly related to the AR-HCF dispersion \cite{kottig_modulational-instability-free_2020}, the ARE wall thickness is a critical parameter for controlling MI in these fibers.

Guided by the theoretical model developed in \cite{kottig_modulational-instability-free_2020} along with our own simulations that reflect our experimental parameters (discussed in detail below), we choose two different AR-HCF geometries: (i) for a relatively high MI gain, an ARE wall thickness $T=300$ nm and a 30 \textmu{}m core diameter (GLOphotonics K9005B2) with a length of 33 cm that we designate \fa; and (ii) to suppress MI, an ARE wall thickness $T=820$ nm and a core diameter of 41 \textmu{}m (GLOphotonics L1910) with a length of 66 cm that we designate \fb.  

\begin{figure}[ht]
\centering
\includegraphics[width=\linewidth]{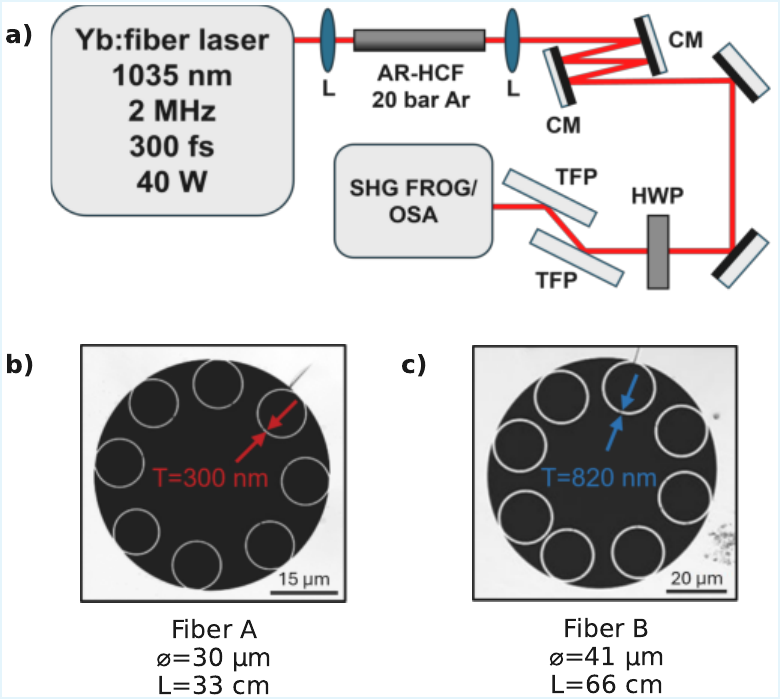}
\caption{a) Experimental setup. The output of a Yb-doped fiber laser oscillator/amplifier system is focused with a lens (L) into an anti-resonant hollow-core fiber  filled with $\approx 20$ bar argon. The output is then recollimated and compressed with chirped mirrors (CM). A half wave plate (HWP) and a pair of thin film polarizers (TFP) act as power control before two diagnostics: second harmonic generation (SHG) FROG and an optical spectrum analyzer (OSA). b), c) Fiber cross section images of \fa{} and \fb, respectively, where $T$ is the anti-resonant element (ARE) wall thickness. }
\label{fig:setup}
\end{figure}

The experimental setup is shown in Fig. \ref{fig:setup}a). Pump pulses at 1035 nm with $\sim$300 fs pulse duration are generated by a home-built Yb-doped fiber laser system operating at 40 W average power output with a repetition rate of 2 MHz. A half-wave plate and polarizing plate beamsplitter (not shown) are used for power control before the laser output is focussed with a plano-convex lens (L) into a gas cell containing one of the spectral broadening AR-HCFs. Microscope images of each fiber end facet are shown in Fig. \ref{fig:setup}b) and Fig. \ref{fig:setup}c). In both cases the gas cell is filled with 20 bar of argon. Total transmission through the fibers is measured to be 88\% for \fa{} and 94\% for \fb. After the fiber, the beam is recollimated with another plano-convex lens and compressed by 12 reflections off chirped mirrors with nominal -150 fs$^2$ group delay dispersion (Ultrafast Innovations PC1611). A second harmonic generation frequency resolved optical gating (FROG) device and an optical spectrum analyzer (OSA) are employed to characterize both the input and compressed output pulses.

\begin{figure}[ht]
\centering
\includegraphics[width=\linewidth]{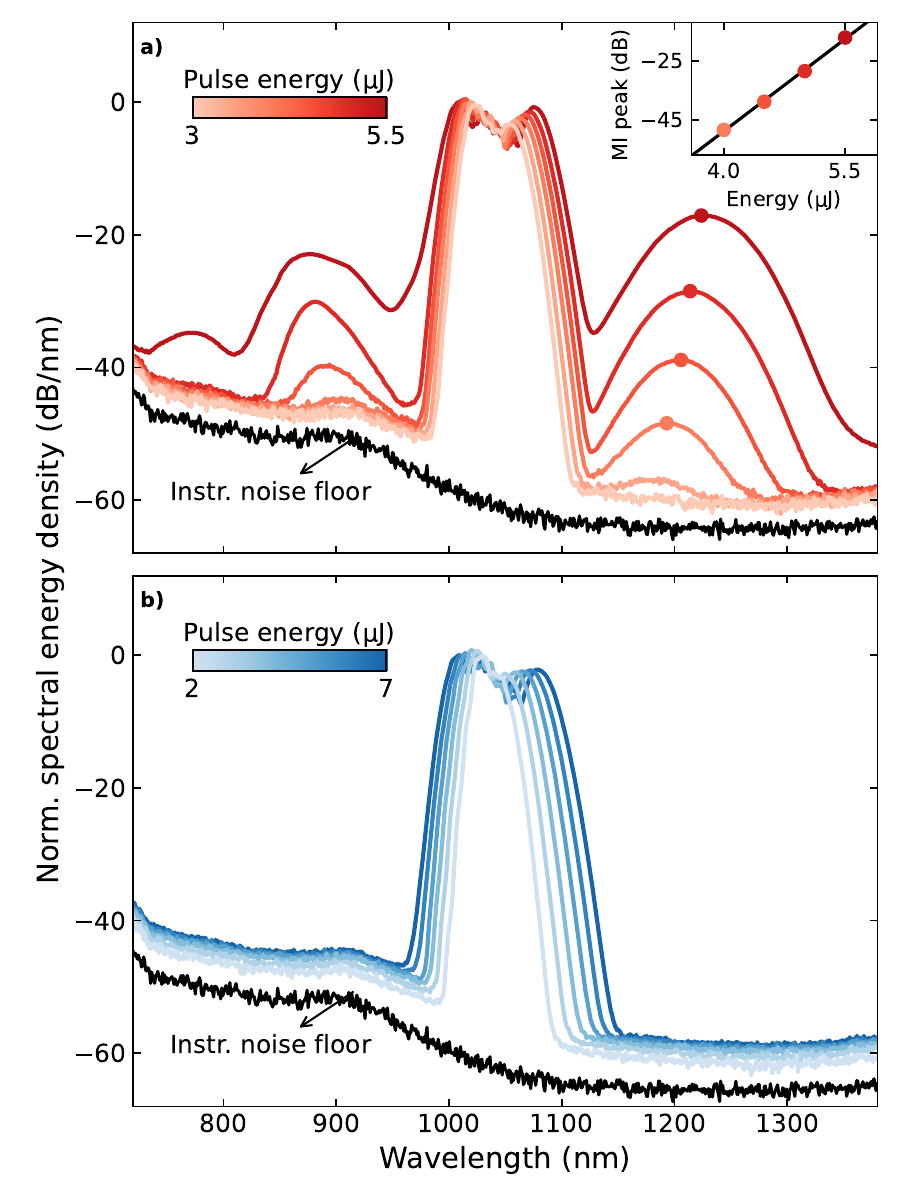}
\caption{Pulse energy scaling of spectral broadening in each AR-HCF. a) \fa{} with a capillary wall thickness of 300~nm showing clear evidence of MI sidebands near 900 and 1200~nm. The inset shows a linear fit to the peak values (dB scale) of the MI sideband near 1200~nm, plotted versus the input pulse energy (linear scale), indicating that the gain is exponential. b) \fb{} with a capillary wall thickness of 820~nm displaying no MI sidebands.}
\label{fig:exp_spectra}
\end{figure}

The output spectra for \fa{} and \fb{}  are shown in Fig.~\ref{fig:exp_spectra}a) and b), respectively, over a range of input pulse energies. Note the ranges are different between the two fibers due to the different core diameters and fiber lengths, and are chosen to demonstrate sufficient spectral broadening to support \textless30~fs full-width-at-half maximum (FWHM) pulse durations. For \fa, MI sidebands begin to appear at 3.5~\textmu{}J and grow rapidly for modest increases in pulse energy. The inset in Fig. \ref{fig:exp_spectra}a) shows the expected exponential gain of MI peak near 1200~nm as a function of the input pulse energy. In contrast, we see no evidence of MI sidebands in \fb{} above the measurement noise floor, for input pulse energies up to 7~\textmu{}J.

\begin{figure}[ht]
\centering
\includegraphics[width=\linewidth]{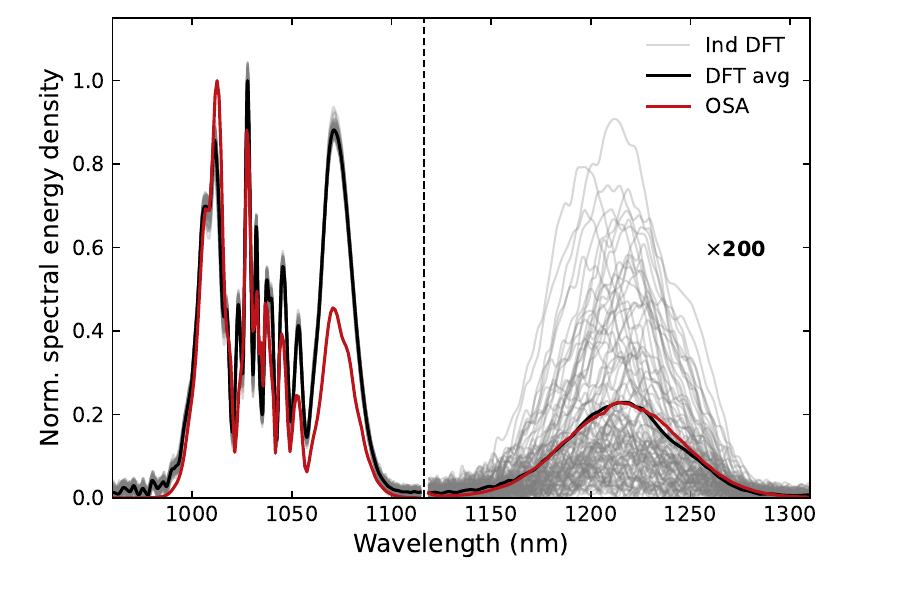}
\caption{Experimental single shot spectra of \fa{} using dispersive Fourier transform (DFT) spectroscopy. Individual spectra are shown in gray while the ensemble average is shown in black. An independently measured average spectrum from an optical spectrum analyzer (OSA) is shown in red. The wavelength axis is calibrated in two separate regions denoted by the black vertical dashed line (see text for details).The region from 1120-1300~nm is multiplied by 200 to show detail of the sideband.}
\label{fig:dft}
\end{figure}

To further characterize the MI sidebands, single-shot spectra of \fa{} at 5 \textmu{}J are measured using a dispersive Fourier transform (DFT) spectrometer \cite{tong_fibre_1997}. In our implementation, 5~km of single mode fiber (Nufern 1060XP) is used as the dispersive element followed by detection with a fast ($\sim$2.5 GHz) photodiode and oscilloscope. Figure \ref{fig:dft} shows 100 individual single-shot spectra in gray, with the ensemble average in black. An independently measured OSA spectrum is shown in red, displaying good agreement with the averaged ensemble in both the main pulse and sideband. Only the MI sideband near 1200 nm is measured due to the single mode fiber loss and poor photodetector sensitivity near 900 nm. The long wavelength side of the OSA spectrum (right of the vertical black dashed line) is multiplied by a factor of 200 and the DFT spectra are scaled to overlap with the spectrum taken by the OSA. The pulse-to-pulse fluctuations observed here confirm the unstable nature of the sidebands, which lead directly to the accompanying peak pulse power variations, as discussed below.

\begin{figure}[ht]
\centering
\includegraphics[width=\linewidth]{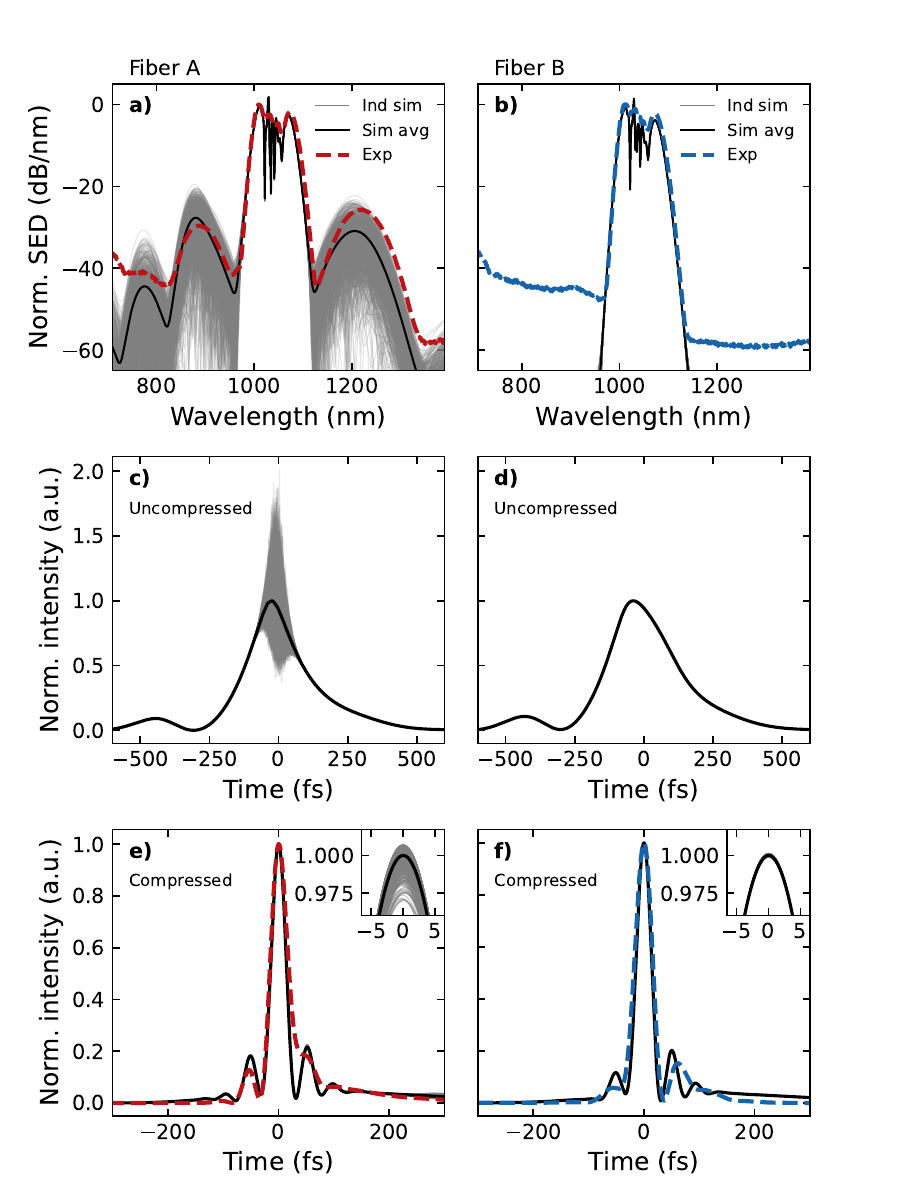}
\caption{Pulse propagation simulations with experimental measurements for T=300 nm (\fa, left column) and T=820 nm (\fb, right column) wall thickness AR-HCFs. One thousand simulations were conducted for each fiber and are shown in gray. The average of each set of simulations is shown in black. Experimental results (when available) are shown in coloured dashed lines. a) and b) output spectra; c) and d) output temporal intensity; e) and f) temporal intensity after numerical compression along with experimental FROG retrievals. Insets show zoomed-in views of the peak fluctuations.}
\label{fig:sims}
\end{figure}

To support these experimental results, we perform numerical simulations using the Luna.jl package \cite{brahms_lunajl_2025}, and we simplify the model to use a single mode in a linear polarization state. We use the Zeisberger and Schmidt model of dispersion for the AR-HCFs \cite{zeisberger_analytic_2017}, and ignore propagation losses under the assumption that the loss bands are sufficiently spectrally distant from the pump wavelength. Input pulses to the AR-HCFs are derived from experimentally measured FROG retrievals of our laser output, having approximately 300~fs FWHM pulse duration. The simulation parameters are chosen to match experimental conditions, with the exception of the pulse energy which was adjusted to overlap the experimentally measured spectra (simulated energies are a factor of 1.2 less for \fa{} and 1.4 less for \fb{}), which we attribute to experimental uncertainties. For each AR-HCF, we run 1000 simulations and include random one-photon-per-mode shot noise \cite{dudley_coherence_2002} to seed MI. The resulting spectra (top), uncompressed (middle) and compressed (bottom) temporal intensities are shown in Fig. \ref{fig:sims} for \fa{} (left column, 4.18~\textmu{}J) and \fb{} (right column, 4.23~\textmu{}J). Individual simulations are shown in gray with ensemble averages in black. Experimentally measured spectra and temporal intensity FROG retrievals are shown in colored dashed lines where available.

First, discussing \fa{} in Fig.~\ref{fig:sims}a), we see excellent agreement between the simulation average and experiment in the SPM portion of the spectrum and good agreement in the MI sidebands at 900 and 1200~nm. The simulated ensemble average are comparable to the measured OSA spectrum and the individual simulations show there are large pulse-to-pulse fluctuations in the sidebands, in agreement with our DFT measurements. In the time domain shown in Fig.~\ref{fig:sims}c), these fluctuations manifest as significant modulations in the peak of the pulse, indicating the onset of pulse break-up. Subsequent numerical pulse compression is achieved with nine reflections off chirped mirrors (UFI PC1611), as shown in Fig~\ref{fig:sims}e). We note this is different than the 12 reflections used in our experiment, which we attribute to unaccounted dispersion in the setup as well as discrepancies in the dispersion model. The simulated FWHM pulse duration is 30~fs and is comparable to the experimentally derived 34~fs from the FROG measurement. A zoomed-in view of the compressed pulse peak (simulations), displayed in the inset, shows peak power fluctuations with a standard deviation of 0.6\%.



In contrast, MI sidebands are notably absent for \fb{} shown in Fig.~\ref{fig:sims}b). The spectrum exhibits negligible variation across individual simulations and shows excellent agreement with the experimentally measured (averaged) spectrum. Similarly, the uncompressed pulse in the time domain, shown in Fig.~\ref{fig:sims}d), exhibits no signs of pulse-to-pulse fluctuations. After numerical compression with 10 chirped mirror reflections, the FWHM duration in Fig.~\ref{fig:sims}f) is similar to \fa{} (30~fs in simulation and 34~fs as retrieved from a experimental FROG measurement) but with significantly smaller peak power fluctuations of 0.04\% (a factor of 15 smaller compared to \fa).

\begin{figure}[ht]
    \centering
    \includegraphics[width=\linewidth]{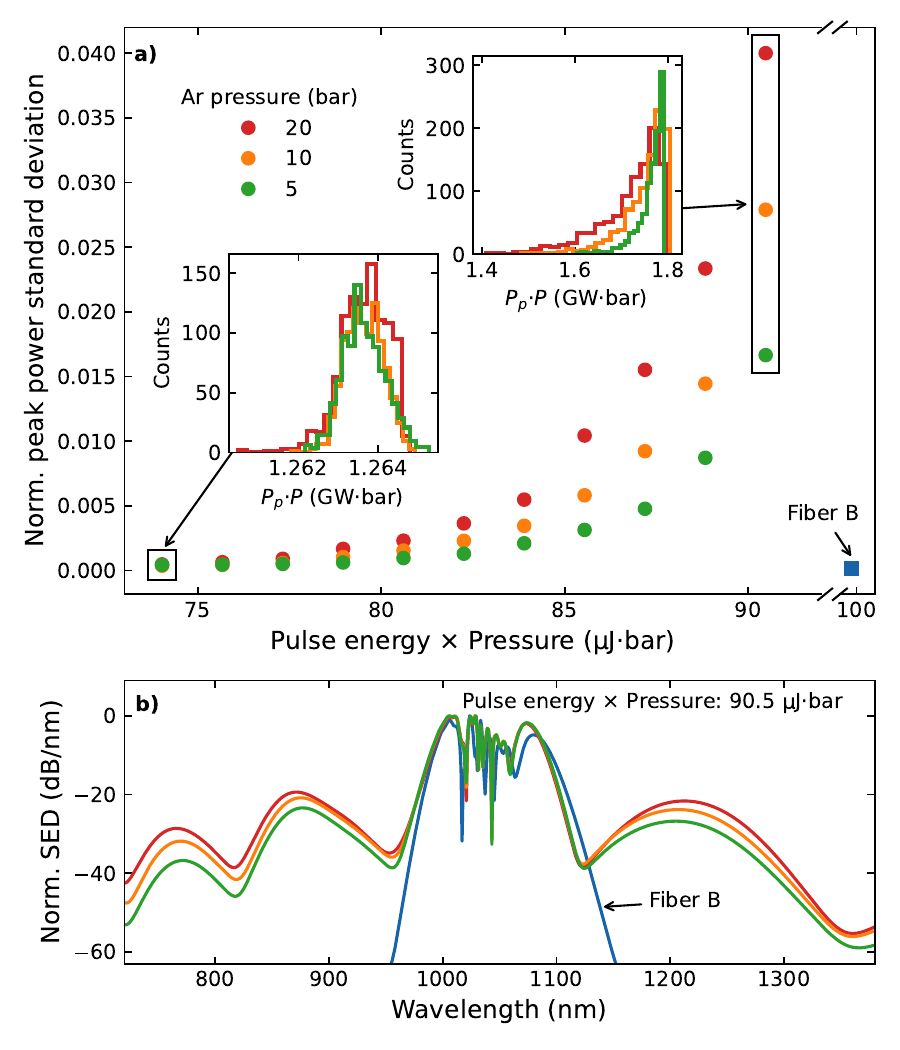}
    \caption{a) Standard deviation of the peak power distribution for 1000 simulations as a function of the product of input pulse energy and argon pressure in \fa. Insets show the peak power ($P_p$) distributions normalized to pressure ($P$) for the points framed by black rectangles. b) Simulated normalized average spectra corresponding to the rightmost points in a). Also shown in blue is a spectrum for \fb{} under comparable broadening conditions with corresponding peak power standard deviation represented by the blue square in a).}
    \label{fig:pk_pow_noise}
\end{figure}

To study the power scaling effects of MI on the compressed pulses, we expand our simulation parameters for \fa{} to pulse energies that approximate the energy range in Fig.~\ref{fig:exp_spectra}a). In addition, we also adjusted argon pressures to 10 and 5~bar, with pulse energies increased by a factor of two and four, respectively, for a comparable nonlinearity between gas pressures. Figure~\ref{fig:pk_pow_noise}a) shows the standard deviation (1000 samples per data point) of the compressed pulse peak power as a function of the product of input pulse energy times pressure. The number of chirped mirror reflections (between 6-8 reflections) is chosen to minimize the pulse duration for each data point. The first observation is an exponential-like increase in the peak power noise as pulse energy is increased, mirroring the growth of the MI sidebands. We also note an increasing noise level with increasing pressure as the latter modifies MI gain via changes in dispersion. In Fig.~\ref{fig:pk_pow_noise}b), we illustrate a similar trend in the MI sideband power versus pressure, and show that the central SPM portion of the spectrum remains constant for a fixed pulse energy-pressure product. We note that \fb{} shows no significant peak power fluctuations at this level of spectral broadening at 20~bar as shown by the blue square in Fig.~\ref{fig:pk_pow_noise}a). We find that scaling at this pressure to a spectrum that supports 20-fs is possible while keeping the peak power fluctuations to the same level as the leftmost point in Fig.~\ref{fig:pk_pow_noise}a). We do not explore this regime in detail as experimental confirmation of the simulations is currently beyond our system capabilities.

The insets in Fig.~\ref{fig:pk_pow_noise}a) show histograms of peak power ($P_p$) distributions from the 1000 simulations again multiplied by pressure ($P$) for the data points framed by rectangles. In contrast, at high energies (right inset), the distributions become skewed to higher peak powers with a long lower peak power tail and have a decidedly non-Gaussian distribution. Such a significant peak power spread has important implications when utilizing the compressed pulses to drive subsequent nonlinear processes as this peak power noise will be directly scaled up by the order of the nonlinear process itself. 

In conclusion, we experimentally verify that controlling AR-HCF capillary wall thickness is essential for suppression of MI and find that this step minimizes pulse peak power fluctuations in SPM-based pulse compression schemes. A T=300~nm ARE wall thickness demonstrates strong MI gain and significant peak power noise, while a T=820~nm ARE wall thickness remains MI free and correspondingly has lower noise up to bandwidths supporting sub-30~fs pulses. Our simulations display excellent agreement with experimentally measured parameters and indicate significant peak power fluctuations in the compressed pulses. Pulse energy and pressure scaling of the peak power fluctuations reveal both exponential-like growth and the emergence of strongly skewed distributions in AR-HCfs supporting MI, a key consideration when employing these compressed pulses to drive subsequent nonlinear processes. Through proper AR-HCF design, we show MI gain can be substantially reduced to enable production of stable ultrafast laser pulses.

\begin{backmatter}
\bmsection{Funding} This research  undertaken thanks in part to funding from the Max Planck-UBC-UTokyo Centre for Quantum Materials, and the Canada First Excellence Research Fund in Quantum Materials and Future Technologies Program. This project is also funded by the Gordon and Betty Moore Foundation's EPiQS Initiative, Grant No. GBMF4779 to  D.J.J.; the Natural Sciences and Engineering Research Council of Canada (NSERC) Discovery Grants program the Canada Foundation for Innovation (CFI); the British Columbia Knowledge Development Fund (BCKDF).

\bmsection{Acknowledgment}We acknowledge helpful discussions with Francesco Tani.

\smallskip

\bmsection{Disclosures} The authors declare no conflicts of interest.

\bmsection{Data availability} Data underlying the results presented in this paper are not publicly available at this time but may be obtained from the authors upon reasonable request.

\bigskip

\end{backmatter}

\bigskip

\bibliography{main}

\bibliographyfullrefs{main}

\ifthenelse{\equal{\journalref}{aop}}{%
\section*{Author Biographies}
\begingroup
\setlength\intextsep{0pt}
\begin{minipage}[t][6.3cm][t]{1.0\textwidth} 
  \begin{wrapfigure}{L}{0.25\textwidth}
    \includegraphics[width=0.25\textwidth]{john_smith.eps}
  \end{wrapfigure}
  \noindent
  {\bfseries John Smith} received his BSc (Mathematics) in 2000 from The University of Maryland. His research interests include lasers and optics.
\end{minipage}
\begin{minipage}{1.0\textwidth}
  \begin{wrapfigure}{L}{0.25\textwidth}
    \includegraphics[width=0.25\textwidth]{alice_smith.eps}
  \end{wrapfigure}
  \noindent
  {\bfseries Alice Smith} also received her BSc (Mathematics) in 2000 from The University of Maryland. Her research interests also include lasers and optics.
\end{minipage}
\endgroup
}{}

\end{document}